# VR-BASED BLOCKCHAIN-ENABLED DATA VISUALIZATION FRAMEWORK FOR MANUFACTURING INDUSTRY

**Nitol Saha**
University of South Carolina
Columbia, SC

**Philip Samaha**
University of South Carolina
Columbia, SC

**Ramy Harik**
University of South Carolina
Columbia, SC

**ABSTRACT**

*This research proposes a blockchain-based data visualization framework integrated with VR to get manufacturing insights. This framework is implemented at the testbed of the Future Factories Lab at the University of South Carolina. The proposed system aims to enhance understanding, analysis, and decision-making processes by immersing users in a VR environment where complex manufacturing data stored using blockchain is translated into intuitive and interactive representations. The project focuses on two main components: blockchain and VR. Hyperledger Fabric is employed to establish a blockchain network, recording data from the Future Factories testbed. This system captures information from various sources, such as potentiometers on robot grippers to measure grip positioning, load cells to gauge pressure, emergency stop buttons, temperature, speed, and vibration sensors on the conveyors. Whenever predefined conditions are met, pertinent data, including sensor ID, timestamp, value, cause, and importance, is securely recorded in the blockchain, signaling the occurrence of a defect within the cell. Data retrieved from the blockchain system is accessed through 'GET' API requests. A VR application is developed using a cross-platform Unity game engine to visualize the data retrieved from the blockchain database. Meta Quest 3 is used as the targeted Head Mounted VR device. The VR application has two C# scripts: one script to query blockchain data using 'GET' API calls and another script converts the JSON object to text data to visualize in the VR system. The proposed system leverages blockchain technology and VR visualization to deliver immersive, actionable insights using secure data transmission. By embracing the proposed framework, manufacturers can unlock new potential for efficiency, sustainability, and resilience in today's increasingly complex and interconnected manufacturing workplace.*

## 1. INTRODUCTION

In modern manufacturing, the utilization of data-driven technologies has become imperative for enhancing operational efficiency and improving product quality. However, this increasing reliance on data also introduces cybersecurity threats. Manufacturing facilities are now vulnerable to cyberattacks that can jeopardize the integrity of critical production data and pose threats to product quality and overall operational stability.

Traditional centralized data storage methods are generally used in manufacturing which is susceptible to cyber threats due to their single point of failure. In this context, blockchain technology emerges as a promising solution. Blockchain's decentralized architecture and cryptographic security features offer a robust framework for the protection and integrity of manufacturing data. By leveraging blockchain technology, manufacturing industries can enhance data security, enable secure data sharing, and establish transparency and traceability throughout the manufacturing lifecycle.

Blockchain technology initially found its footing in financial applications [1]. The enhanced security features in blockchain drew the attention of researchers exploring its potential in different domains. It was within this context that Ethereum emerged, offering a public blockchain framework, with the added capability of accommodating diverse data structures [2]. However, the security of public blockchain systems heavily relies on mining, a resource-intensive process, which is essential for validating network transactions and maintaining security, alongside the decentralized nature of blockchain. In practical industry settings, the substantial resource overheads of these blockchain architectures pose challenges, especially considering the need for equivalent levels of security [3].

Following this, the concept of private blockchain architecture has emerged to address industry-specific needs, especially for the manufacturing industry [4]. Unlike public blockchain models, private blockchains prioritize security over public accessibility. Instead, it adopts a permissioned approach, ensuring robust security standards by controlling access through permissions. In the private blockchain, the initiating organization retains the authority to grant special permissions to other entities seeking to join the network [5]. For this reason, numerous platforms have been developed to accommodate the diverse requirements of different industries, for instance, Hyperledger Fabric, Multichain, Corda, and Stellar.

Furthermore, the integration of emerging technologies like Virtual Reality (VR) into manufacturing processes presents new opportunities for operational optimization and quality control. VR allows users to interact with digital representations of manufacturing environments, facilitating real-time process monitoring, interactive maintenance, training, and quality assurance [6]. However, the adoption of VR introduces its own set of data security challenges, particularly concerning the integrity and confidentiality of virtual manufacturing data.

The synergy between blockchain and VR technologies holds immense potential for manufacturing operations. blockchain's inherent security features can address the vulnerabilities associated with VR-enabled manufacturing systems, ensuring the trustworthiness and reliability of data.

By exploring the intersection of blockchain and VR, this paper proposes a comprehensive framework for secure and efficient immersive data visualization. This framework aims to mitigate cybersecurity risks in VR systems and enhance data integrity.

The objective is to establish a unified system that integrates all nodes into a single network for seamless data sharing and visibility during data visualization in VR systems. By achieving this, it becomes possible to monitor and trace anomalies occurring across all processes of manufacturing efficiently. This includes the ability to pinpoint where anomalies occur, what specific events happened, and extract valuable insights into the underlying causes. By securely storing this data in a secured

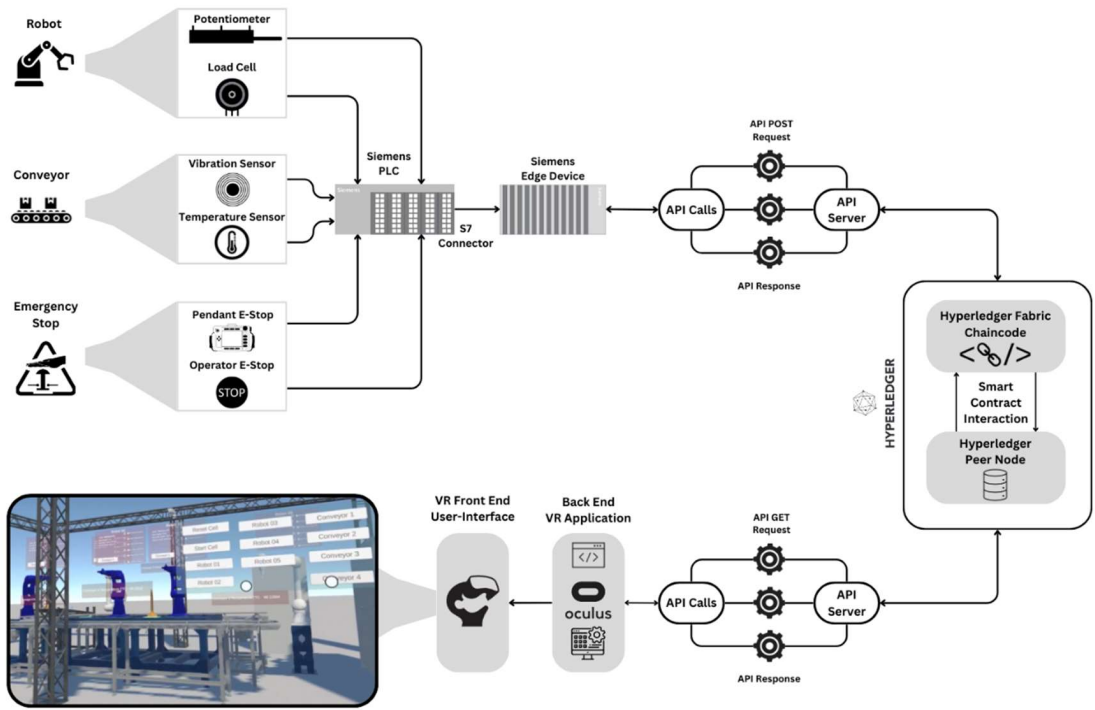

**FIGURE 1:** PROPOSED FRAMEWORK

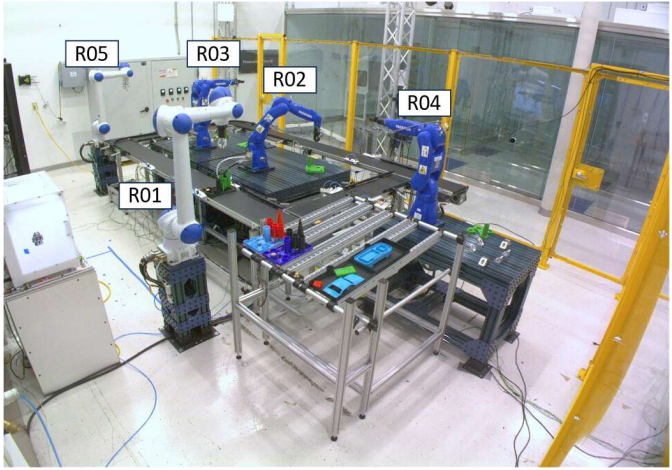

**FIGURE 2:** FUTURE FACTORIES TESTBED

database it mitigates the chances of malicious behavior during the manufacturing operations.

The subsequent sections will delve into the methodology, implementation strategies, and preliminary results of the proposed framework in detail.

## 2. METHODOLOGY & IMPLEMENTATION

In this section, the detailed methodology of the proposed framework (Figure 1) and the implementation strategies of the framework within the Future Factories testbed (Figure 2) at the University of South Carolina (USC) are discussed. The implementation covers several layers of the technological assets, including manufacturing equipment, edge computing, blockchain integration, communication protocols, and immersive visualization techniques. Each layer is working together to build a comprehensive system for blockchain-enabled VR-based manufacturing data visualization framework.

### 2.1 Equipment Layer

The initial asset category is within the equipment layer, including actuators, transducers, and controllers. The Future Factories testbed features five robotic arms that primarily handle product manipulation. Two Yaskawa HC10s (R01 and R05) manage material handling, while three GP8s (R02, R03, and R04) perform collaborative assembly and disassembly of rocket components (Figure 2). Additionally, there is a conveyor system comprising four interconnected for the internal movement of rocket parts between assembly stations [7].

Different sensors were installed to collect data from the system. These sensors are integrated with a Siemens S7-1500 Programmable Logic Controller (PLC) which carries out the logic of the assembly process. The PLC's multiple I/O modules are used to connect analog and digital sensors. Profinet communication protocol is used to exchange data with robot controllers and conveyors' drives. The following sensors are used in the testbed:

- Potentiometer on R01, R02, R03, and R04
- Load Cell on R01, R02, R03, and R04
- Pressure Gauge on R05
- Temperature Sensors on Conveyors 1, 2, 3, and 4 actuators and controllers
- Speed sensors on Conveyors 1, 2, 3, and 4
- Vibration Sensors on Conveyors 1, 2, 3, and 4
- Emergency Stop Button on R01, R02, R03, R04, and R05
- Emergency Stop Button on the Human Machine Interface (HMI)
- Emergency Stop Button on the main Control Panel

The potentiometers on the robots' grippers provide insight into the degree of closure and opening of the gripper. By establishing a predefined threshold for specific operations, we can determine whether the gripper has reached the required level of closure or opening. Load cells and pressure gauges integrated into the robots' grippers measure the pressure applied to handled objects. High-pressure readings may signify improper gripper behavior. Temperature sensors of conveyors also serve as a process monitoring indicator. Exceeding preset thresholds for these parameters signals potential malfunctions in the conveyors. For example, the speed of the conveyors is related to the temperature of the conveyors. When the conveyors operate continuously for an extended duration, they may reach high temperatures, prompting them to enter into safety mode, and their operating speed is reduced. Additionally, emergency stop halts the entire manufacturing process, serving as a clear indication that an event has occurred. These event data are stored for further analysis and visualization.

Furthermore, a demonstrator was developed to emulate a supply chain node that monitors the environmental conditions of a container during product transportation (Figure 3). The demonstrator has Peltier elements with heat exchangers and fans so that the demonstrator can dissipate heat from the container.

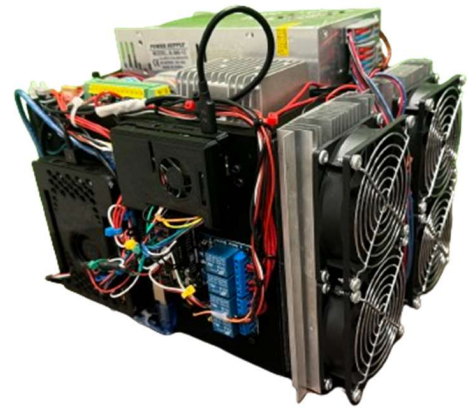

**FIGURE 3:** SUPPLY CHAIN DEMONSTRATOR

The container has various sensors to monitor environmental parameters, including:

- Temperature Sensor
- Humidity Sensor
- Gyroscope Sensor
- GPS Tracker

Predefined thresholds are established for these parameters, to record anomalies within the blockchain database when any sensor data exceeds the set threshold. The recorded data includes details of the nature of the anomaly, its location, sensor readings at the time of occurrence, as well as the corresponding shipment ID and timestamp, which ensures traceability throughout the supply chain process.

## 2.2 Edge Layer

Another important piece of equipment in the framework is the edge device. The IPC227E is used as the edge device for this framework that collects data from different sensors of the Future Factories testbed. It has a 240GB storage drive and 8GB of memory, and it runs on an Intel Celeron N2930 processor. It has two ethernet ports: one connects to the PLC system, and the other links to the USC network. This network connection lets users access the IPC using the Siemens Industrial PC (IPC) Management System on the USC network. It is used to connect to testbed equipment and gather data in real-time. Different applications can be deployed to the edge device for data visualization, analysis, and integration with cloud services. For This framework, this edge device is used to collect sensor data from the testbed to analyze it for sending important data to a blockchain database using API POST calls. This helps identify defects and only relevant data is stored in the blockchain to avoid filling up the database with extra information since data in the blockchain cannot be deleted.

## 2.3 Blockchain Layer

All the mentioned data are stored securely in the blockchain ledger, which prevents anyone from altering or deleting it. This is important because these data directly affect product quality. Using traditional databases poses a risk as they can be manipulated to conceal anomalies in the manufacturing processes. With the blockchain system, these modifications become unfeasible. As a result, any anomalies that occur can be traced back to their origin. The structure of the Future Factories blockchain data is shown in Figure 4.

The figure shows a sample data block of the load cell located at R02. The first line denotes the sensor by its ID. The second line describes the type of fault detected, which helps in categorizing when querying for information for the database. The specific sensor value at the time of fault is also included in the data block. The importance of the anomaly is noted: If the anomaly disrupts the process, it triggers an alert, but if the anomaly allows the process to continue, it generates a warning for operators.

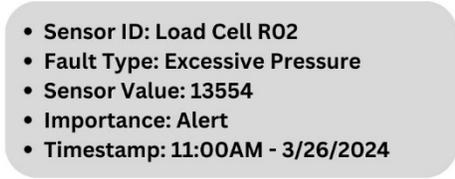

- Sensor ID: Load Cell R02
- Fault Type: Excessive Pressure
- Sensor Value: 13554
- Importance: Alert
- Timestamp: 11:00AM - 3/26/2024

**FIGURE 4:** BLOCKCHAIN DATA STRUCTURE

For this proposed framework, Hyperledger Fabric is used as the blockchain platform. It offers a range of features and is known for its strong security measures. Leveraging decentralized blockchain architecture, it uses sophisticated algorithms to address security concerns in distributed networks [8]. Moreover, this platform supports the flexibility to accommodate various consensus algorithms, each serving distinct purposes while ensuring the required level of security. As an example, the RAFT consensus algorithm serves to mitigate challenges arising from hardware or software failures within a network, ensuring continuity and preserving transaction sequencing even in the event of node disruptions or software anomalies. However, it is important to note that while proficient in addressing common operational issues, RAFT does not inherently provide defenses against more sophisticated forms of corruption, such as cyber-attacks. To address such concerns, Hyperledger Fabric offers the capability to accommodate more sophisticated consensus algorithms designed to counter these challenges. One notable example is the Practical Byzantine Fault Tolerance (PBFT) algorithm, known for its role in fortifying networks against cyber-attacks and malicious activities [9]. However, for this project, a relatively straightforward consensus algorithm was selected for demonstration purposes. The chosen consensus algorithm for this project was "KAFKA", a modified version of RAFT designed with optimized features to address decentralized network challenges.

In addition, Hyperledger Fabric allows the creation of specific channels and chaincodes for individual use cases. These channels play a critical role in addressing a key challenge in

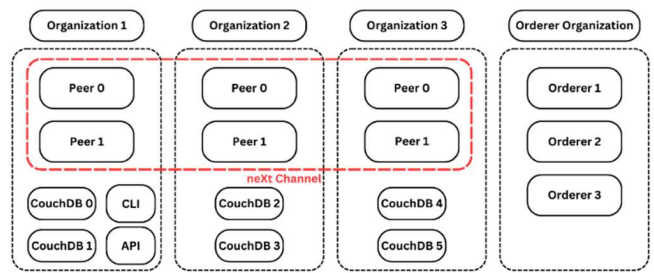

**FIGURE 5:** BLOCKCHAIN NETWORK ARCHITECTURE

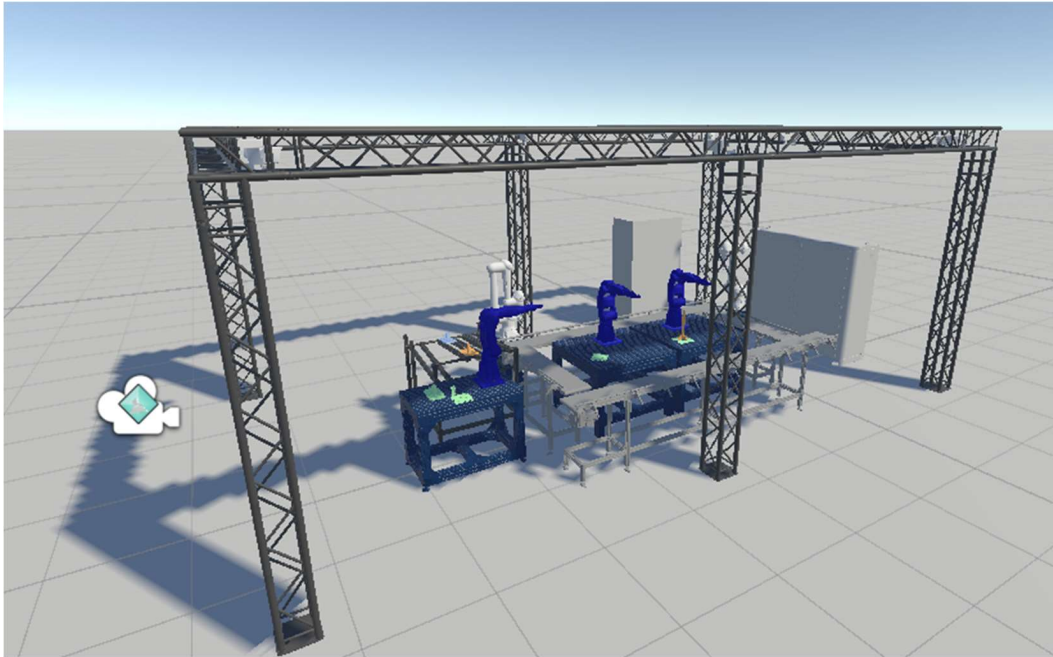
**FIGURE 6:** DIGITAL TWIN REPRESENTATION OF THE FUTURE FACTORIES TESTBED IN UNITY

decentralized systems. For example, when two organizations collaborate on a project within a blockchain network, they expose their data to each other. However, if they are competitors on another project, full data visibility can impede collaboration. Hyperledger Fabric addresses this by introducing channels, which enable the establishment of distinct communication pathways. Nodes can selectively join these channels based on their relevance to the project. This approach allows collaborating organizations to share only relevant data for a specific project using a dedicated channel. Meanwhile, different projects involving other collaborators can utilize separate channels for data sharing. Furthermore, chaincodes, which are written in common programming languages like Go, Java, JavaScript, or TypeScript, control the logic and operations within each channel to ensure consistency and integrity across the network. These chaincodes handle essential functions for recording and retrieving data in the blockchain ledger.

The proposed blockchain network consists of four organizations, where three serve as nodes and the fourth as the orderer organization. An API (Application Programming Interface) server is deployed to enable interactions with the network from client applications, and a channel is created to include all the peers within the network. This finalized network architecture is illustrated in Figure 5.

### 2.4 Communication Layer

An API (Application Programming Interface) interacts with the blockchain ledger. Using the API server, client applications input data onto the blockchain ledger and query data from it. This API method offers a standardized way to connect with the blockchain network, allowing smooth integration and compatibility across different applications and systems. API calls are made over standard internet communication protocols (TCP/IP).

### 2.5 Immersive Visualization Layer

With the deployment of a blockchain system and its associated applications for data integration, the next step is to visualize this data within a VR platform. To achieve this, a digital twin representation of the Future Factories testbed was developed using the cross-platform Unity Game Engine (Figure 6). Unity is one of the most popular VR development platforms due to its support for various VR Software Development Kits (SDK) and compatibility with major Head Mounted Devices (HMDs). For this project, Meta Quest 3 HMD was selected for the VR visualization. To integrate Meta Quest 3 into Unity, the Meta XR All-in-One SDK (Version 64.0) was utilized. This latest SDK integrates all the essential Interaction and Haptic SDKs together which is needed for seamless integration and immersive user experiences within the VR environment. This setup allows for the effective integration of blockchain data visualization within the VR environment.

This implementation utilizes a C# Unity script to make the API calls to the blockchain-based endpoint that communicates with RESTful API calls to retrieve and process JSON data for a specified shipment as well as individual sensors of the Future Factories lab (Figure 7). The script uses an authentication flow by sending a POST request to the blockchain database with predefined request body parameters. Upon successful authentication, it receives a registration token from the response which is utilized in the subsequent API requests. Using the registration token, this script fetches the defect data related to a specified shipment and also for individual sensors whichever is requested from the blockchain-based API endpoint. This is achieved by sending an authorized GET request with the obtained registration token in the authorization header.

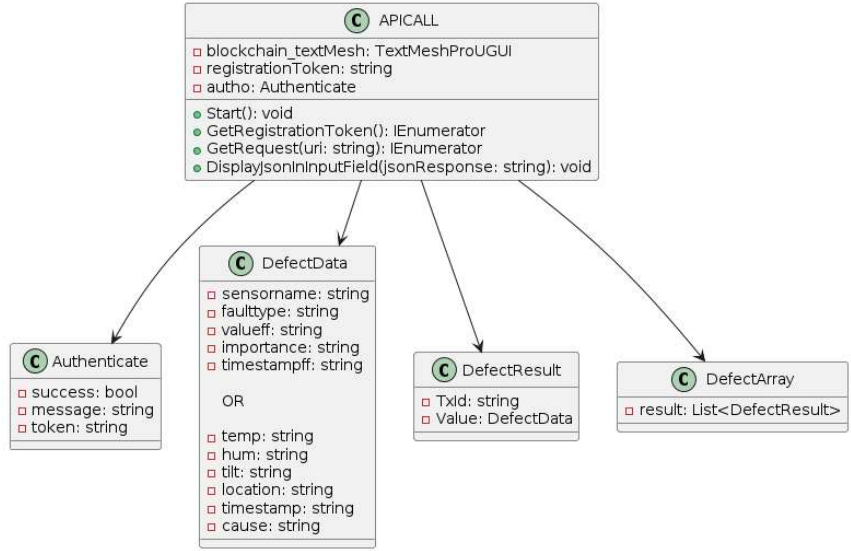

**FIGURE 7:** STRUCTURE OF THE API CALLS

Using another script, the response JSON data is then deserialized into a structured C# class using "Newtonsoft.Json", to display relevant defect information from the blockchain. The retrieved defect data is then appended to the "TextMeshProUGUI" objects within the Unity scene for real-time display and visualization of the blockchain-based defect information.

Furthermore, this VR platform has additional functionalities such as real-time data visualization via MQTT, interaction with diverse assets, and the ability to control robots for digital twinning applications.

Finally, the VR application is deployed as an Android Application on Meta Quest 3, for a standalone and immersive

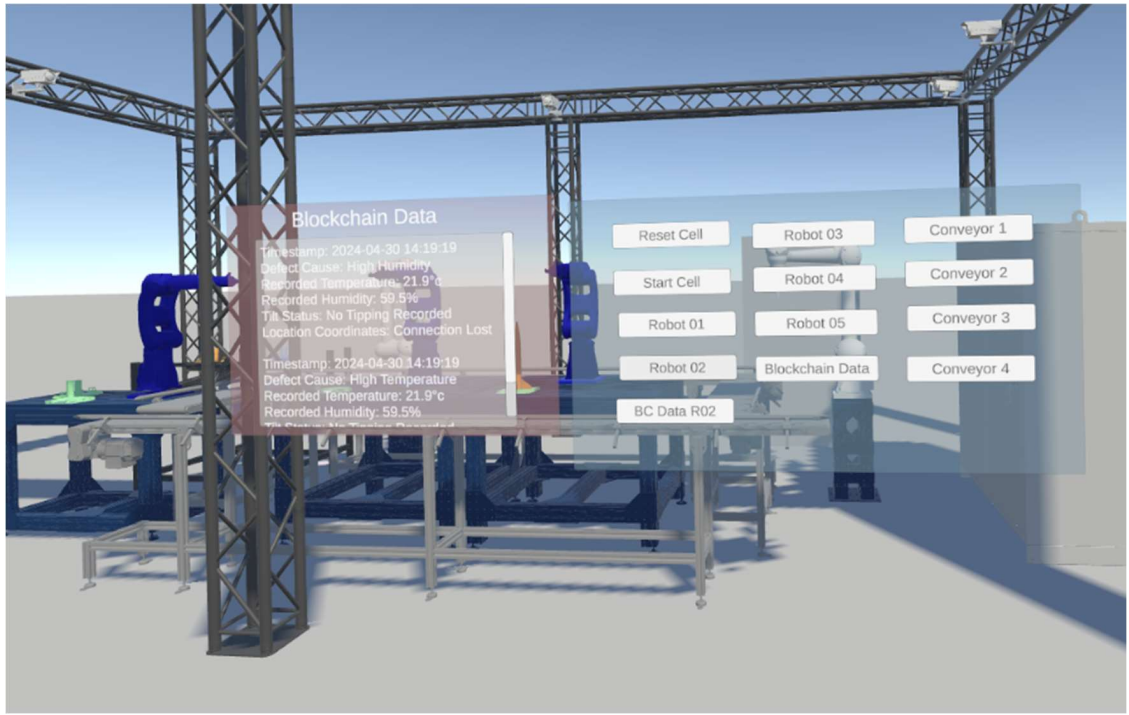

**FIGURE 8:** BLOCKCHAIN-BASED SHIPMENT DATA VISUALIZATION USING DASHBOARD IN VR

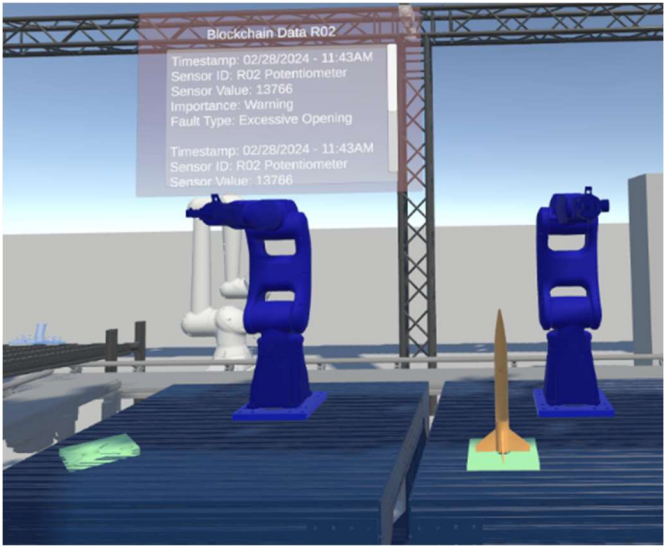

**FIGURE 9:** VISUALIZATION OF FAULTY DATA FOR INDIVIDUAL ASSEST RETRIEVED FROM THE BLOCKCHAIN DATABASE

experience for VR data visualization and interaction of the virtual Future Factories testbed.

## 3 RESULTS AND DISCUSSION

The incorporation of blockchain data visualization into the VR platform utilizing Unity and Meta Quest 3 has produced effective results for the digital twin representation within the Future Factories testbed (Figure 6).

This VR application has several features. Upon entering the immersive VR environment, users can navigate using the teleportation feature provided by the Meta All-in-One SDK. This functionality allows for seamless movement within the virtual space. Additionally, users can utilize the controllers of Meta Quest 3 to grab and interact with objects within the environment for a hands-on experience. The VR application features a dashboard that offers various options for data visualization (Figure 8). Users can access this dashboard in the VR environment and click on specific buttons to retrieve detailed information based on their preferences. For instance, by clicking the 'Blockchain Data' button, users can access information related to defects during shipment of the supply chain demonstrator (Figure 8).

Upon clicking the 'Blockchain Data' button, users are presented with detailed information regarding shipment defects. This includes the time of occurrence, defect cause, tilt status (tilting information retrieved from Gyroscope), and recorded sensor data, all displayed within the 'Blockchain Data Canvas'. This data is retrieved via API calls for real-time visualization of defect-related information. Users can use the controllers of the Meta Quest 3 to scroll and view detailed defect information displayed on the 'Blockchain Data Canvas'. This interaction is facilitated by the Ray interaction feature of the Meta All-in-One SDK.

Another feature of the VR application is the ability to view individual blockchain-based defect data associated with specific assets. For instance, by selecting 'BC Data R02' on the dashboard, users can visualize defect data over Robot 02 (Figure 09). This feature enables users to identify asset defects for further analysis and troubleshooting. Users can view detailed fault information associated with different sensors of the selected asset, including the time of occurrence, sensor ID, sensor value, importance of the fault, and fault type. This comprehensive visualization capability aids in efficient defect identification and management within the virtual environment. The utilization of blockchain technology ensures the authenticity and security of the data against alterations and cyberattacks.

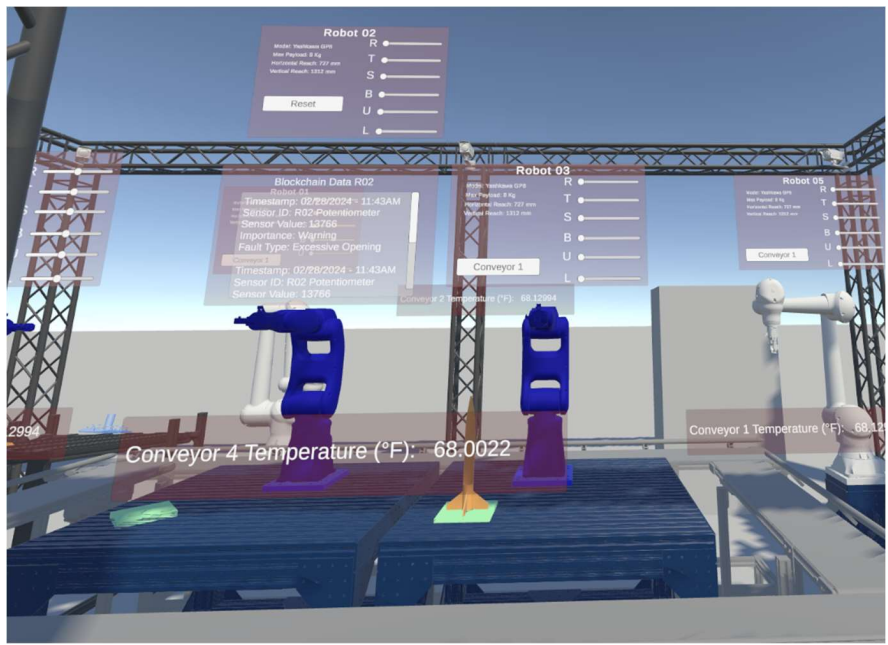

**FIGURE 10:** DIFFERENT CAPABILITIES OF THE VR PLATFORM

The VR platform offers a range of capabilities beyond defect visualization (Figure 10). Users can access real-time sensor data through the dashboard for the monitoring of operational metrics and performance indicators for different assets. Additionally, the platform supports interactive digital twinning applications, such as allowing users to start or stop the real Future Factories testbed directly from the VR interface. Furthermore, users can manipulate robot joints using the VR platform for training and simulation purposes. These integrated functionalities demonstrate the versatility of the developed VR application.

## 4 CONCLUSION

The proposed framework showcases the potential of combining blockchain technology with virtual reality (VR) for secured data visualization. By leveraging blockchain, this framework ensures data transparency and integrity by providing a secure and immutable ledger for manufacturing data while safeguarding against cyberattacks and unauthorized alterations of data. The immersive VR platform developed for the Future Factories testbed enables secured real-time defect visualization and asset monitoring. Additionally, interactive digital twinning capabilities enable simulation and training scenarios of manufacturing operations. However, challenges persist in deploying blockchain networks due to the complexity of implementation and a shortage of skilled developers proficient in implementing blockchain networks for manufacturing applications. For future work, the framework will be extended to address diverse challenges of supply chain management, quality assurance, and predictive maintenance. In conclusion, the successful deployment of the VR application at the Future Factories testbed underscores the transformative potential of VR-based secured data visualization techniques within the manufacturing industry by utilizing blockchain.

## ACKNOWLEDGEMENTS

This work is funded in part by NSF Award 2119654 "RII Track 2 FEC: Enabling Factory to Factory (F2F) Networking for Future Manufacturing," and "Enabling Factory to Factory (F2F) Networking for Future Manufacturing across South Carolina," funded by South Carolina Research Authority. Any opinions, findings, conclusions, or recommendations expressed in this material are those of the author(s) and do not necessarily reflect the views of the sponsors.